\documentclass[twocolumn,aip,jcp,lengthcheck,showpacs,floatfix]{revtex4-1}
\usepackage[]{amsmath}
\usepackage{amssymb}
\usepackage[version=4]{mhchem}
\usepackage{siunitx}
\usepackage{graphicx}
\usepackage{bm}
\usepackage{times,psfrag,subfigure}
\usepackage[]{color}
\usepackage{syntonly}
\usepackage{lmodern}
\makeatletter
\renewcommand*{\@fnsymbol}[1]{\ensuremath{\ifcase#1\or \dagger \or* \else\@ctrerr\fi}}
\makeatother

\begin{document}

\title{Hole-initiated melting process of thin films}

\author{Chenyu Jin}
\altaffiliation[Present address: ]{Uni Bayreuth, Univertit\"atstr. 30, 95447, Bayreuth}
\email{dr.chenyu.jin@gmail.com}
\author{Hans Riegler}
\email{Hans.Riegler@mpikg.mpg.de}
\affiliation{Max Planck Institute of Colloids and Interfaces, Science Park Golm, 14476 Potsdam, Germany}
\date{\today}

\begin{abstract}
We perform numerical and experimental studies on the melting process of thin films initiated by a small hole. 
The presence of a non-trivial capillary surface, namely the liquid/air interface, leads to a few counter-intuitive results:
(1) The melting point is elevated if the film surface is partially wettable, even with a small contact angle. (2) For a film that is finite in size, melting may prefer to start from the outer boundary, rather than a hole inside. (3) More complex melting scenario may arise, including morphology transitions, and the ``de facto'' melting point being a range instead of a single value. These are verified by experiments on melting alkane films between silica and air. 
This work continues a series of investigations on the capillary aspects on melting.
Both our model and analysis approach can be easily generalized to other systems.

\end{abstract}
\maketitle

\section{Introduction}

Thermostability of thin films is of great interest in theory and in practice \cite{huber2015soft}.
For example, it is very important to know when and how an anti-corrosion coating, a filter membrane, or a semi-conductor layer, would collapse under high temperature. 
Here, the thermostability is characterized by the ``de facto''  melting point of the thin film, which, due to the ineligible interfacial free energies, is different from the melting point of the bulk material.

The most used equation for calculating the ``de facto''  melting point of a confined system is the Gibbs-Thomson equation \cite{Pawlow1909a, tammann1920methode, meissner1920, rie1923}.
It predicts that the melting point of a confined system is lower than the bulk melting temperature, with a shift that is inversely proportional to the characteristic length of the confinement.
This theory is deduced from a classical two-phase model: a small solid particle covered by its melt, with one solid-liquid interface. The area of this interface varies during melting, hence the interfacial energy varies and contributes to the total free energy change.
For the thin films, a variation of this model is used, but still only the solid-liquid interface is considered. 

However, in most real-life cases, there will be at least one capillary surface, i.e., the liquid-air interface. 
The prerequisite of the Gibbs-Thomson equation is that the liquid melt covers uniformly a solid particle floating in the air, so that the capillary surface is a spherical shell that doesn't change in shape or area.
In practice, this means all the facets of the floating particle must be ``premelting'' \cite{Frenken1985surfacemelting, Dash1989surfacemelting}, i.e., a thin layer of liquid melt covers the solid facets \emph{before} the temperature reaches the bulk melting point.

A bad news is that ``premelting'' is only confirmed on certain facets of ice and lead (Pb). Experimental study with lead particles shows that the Gibbs-Thomson equation breaks down once a non-premelting facet is involved, namely, the liquid melt wets the facet with a non-zero contact angle \cite{metois1989Pb111}.
This will also happen when the particle is not floating but in contact with a substrate.
Liquid could be pinned at the grain boundaries, or wet the substrate with a non-zero contact angle. The shape of the capillary surface will become non-trivial, so does the interfacial energy. 
For long it has been known that the effects of the capillary surface on melting is complex \cite{maeda1999direct, christenson1995liquid, fretwell1996anomalous, glicksman2006capillary, moerz2012capillary}. But still, this problem is rarely systematically analysed.

Previously, we have observed an unexpected melting scenario of molecularly thin terrace of long-chain alkanes: liquid drops appear at the edge and ``eat'' into the terrace \cite{Lazar2005movingdrop}. 
We have revealed that the emergence of these drops is due to (1) the co-existence of liquid and solid under equilibrium, and (2) the Plateau-Rayleigh instability of the capillary surface \cite{Halim2012bulge}. 
Later we have analysed the melting behaviour of cylindrical aggregates \cite{jin2016island} in gaseous environment based on the same experimental system, 
and have shown a elevated melting point due to non-premelting facets.

In this work, we study the melting behaviour of a thin, planar solid film from a defect in the shape of a cylindrical hole. 
During the melting, the capillary surface may appear in different shapes.
We show that (1) the hole inside the film may not be the preferred starting point of melting; 
(2) if melting starts from the hole, morphology transitions at different instants during the melting will lead to different melting scenarios as well as different ``de facto'' melting points.
Although the simulation is based on our experimental system, the long-chain alkane film between air and silica substrate, the model can be easily generalised to other systems.

\section{Theory and Methods}
\subsection{Quasi-static melting process}
\begin{figure}
    \centering
    \includegraphics[width=\columnwidth]{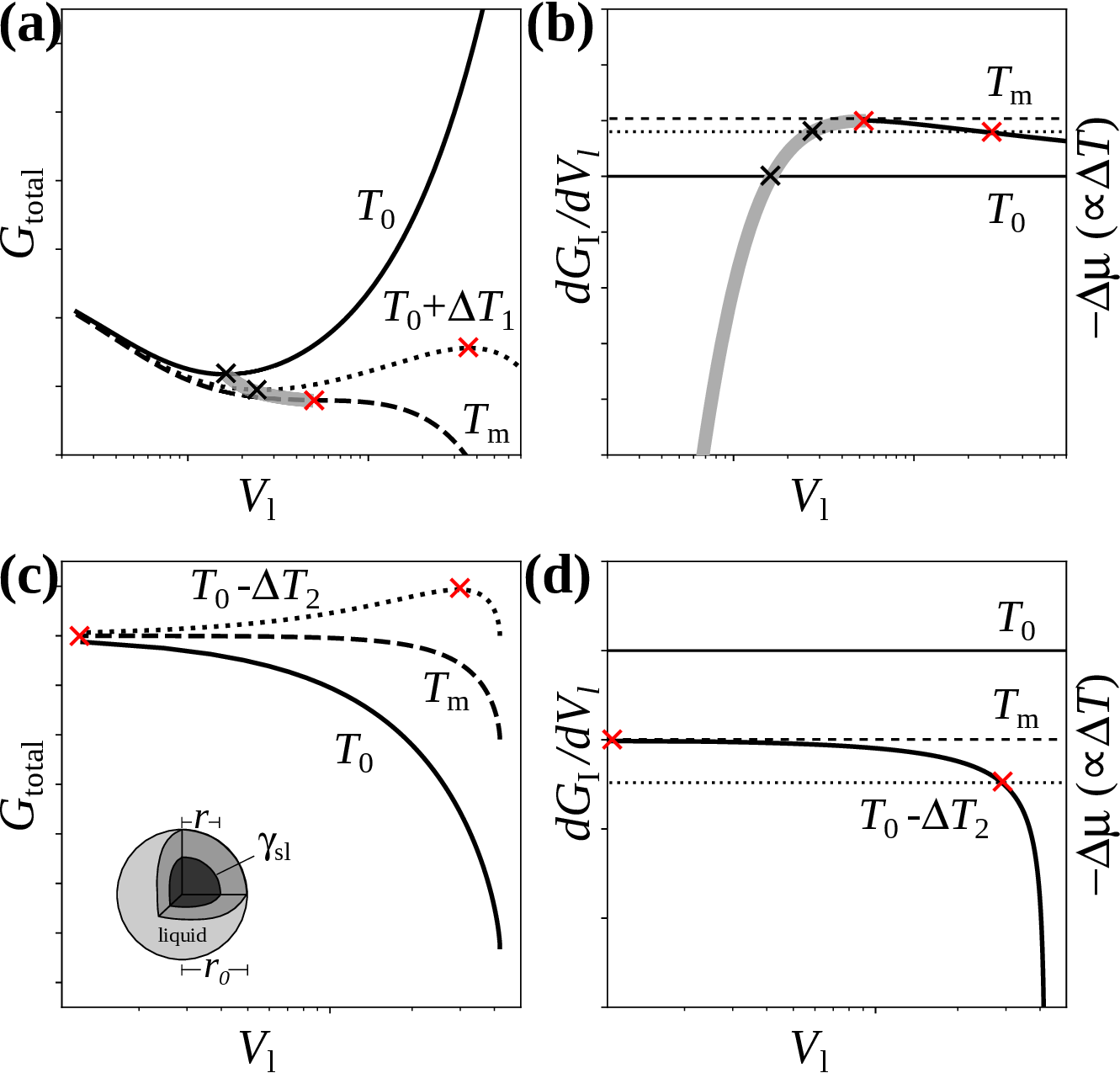}
    \caption{Free energy change during melting of two typical systems. (a) and (b) correspond to an example system: a film melting from a hole of ${r_0}/h=20$. (c) and (d) correspond to the typical Gibbs-Thomson system, namely, a premelting spherical particle with the initial radius $r_0$. }
    \label{fig:thermo}
\end{figure}

We analyze the melting process within a continuum framework. 
\footnote[2]{B. Henrich et al. showed that continuum models are adequate for $\geq 2$ adjacent fluid mono-layers using computer simulation \cite{henrich2008continuum}.}  
Contributions from the line tension \cite{tolman1949effect, berg2010impact}, the disjoining pressure \cite{derjaguin1978}, and the gravity are neglected.
When alkane solid melts, the volume increases typically by $\approx 10\%$ \cite{seyer1944density}. This change in volume can be neglected without affecting our conclusion.
We hence use the volume of the liquid melt $V_l$ to represent the amount of melted material.
Melting is treated as a slow, quasi-static process: only an infinitesimal change of parameter is applied for the status change. Therefore, the capillary surface is always fully relaxed to the minimum of the interfacial energy (area). 

Note that the quasi-static approach is possible in our system because we have observed the co-existence of liquid and solid under (meta-)stable equilibrium \cite{Halim2012bulge, jin2016island}. This is caused by the existence of non-premelting facets and the non-trivial capillary surface, which is not the case for a bulk system or a Gibbs-Thomson system.
Following we analyse the melting process in details.

The system status is determined by the total free energy $G_\text{total}$, which is the sum of the bulk energy $G_\text{B}$ and the interfacial energies $G_\text{I}$. 
Fig.~\ref{fig:thermo}(a) illustrates the total free energy as the function of $V_l$ in an example system.
Three different temperatures are chosen: the bulk melting temperature $T_0$, the ``de facto'' melting temperature $T_m$, and a temperature in between.
Note that at $T_\text{0}$, $G_\text{B}=0$ hence $G_\text{total} =G_\text{I}$ (black curves). 
Each (local) minimum on the energy curve indicates a (meta-)stable equilibrium state, as marked with black crosses.
Red crosses mark the unstable equilibrium states, above which the melt will complete without any energy barrier. 
The grey line connecting the (meta-)stable equilibrium states indicates the quasi-static melting process.
It ends when the system reaches the temperature at which only unstable equilibrium is possible. We call this temperture the ``de facto'' melting temperature $T_\text{m}$ of the system.
\footnote[4]{Sometimes called the ``true'' or the ``critical'' melting temperature.}

In order to decide the scenario of melting as a quasi-static process, we need to find out the equilibrium curve that plot how the system parameters change from the initial (all solid) to the final state (all liquid). 
At equilibrium we have $\partial G_\text{total} / \partial V_l =0$, or
\begin{equation}
  \Delta \mu +\frac{dG_\text{I}}{dV_l} =0,
  \label{eqn:equil}
\end{equation}
where $\Delta \mu$ is the bulk chemical potential difference per volume between liquid and solid under a constant pressure (in this case, 1 bar). 

On the one hand, $\Delta \mu$ only depends on the temperature $T$. More specifically, when the relevant temperature range is small so that a constant melting entropy $\Delta S_\text{fus}$ can be assumed, $\Delta \mu$ at an arbitrary temperature $T$ is calculated as \cite{adkins1983equilTD, Dash2006premeltice}
\begin{equation}
  \Delta \mu \simeq - \Delta S_\text{fus}\cdot (T-T_\text{0}),
  \label{eqn:mu}
\end{equation}
where $T_\text{0}$ is the bulk melting point. That is, $\Delta \mu$ decreases linearly with the temperature $T$.

On the other hand, $dG_\text{I} / dV_l$ does not change with $T$. Rather, it is a function of $V_l$.
The equilibrium curve should hence be plotted with $T$ and $V_l$ as the parameters.
From Eqs.~\eqref{eqn:equil} and \eqref{eqn:mu}, we have the relation
\begin{equation}
  \Delta T = T - T_\text{0} = \frac{1}{\Delta S_\text{fus}} \frac{dG_\text{I}}{dV_l}.
  \label{eqn:T}
\end{equation}

In Fig.~\ref{fig:thermo}(b), $dG_\text{I} / dV_l$ is plotted as a function of $V_l$.
Also, for the three typical temperatures in Fig.~\ref{fig:thermo}(a), we plot the corresponding $-\Delta \mu$, which is proportional to $\Delta T$. They appear as horizontal lines as $\Delta \mu$ does not depend on $V_l$.
Their intersections with the $dG_\text{I} / dV_l$ curve indicate the equilibrium states, as marked by crosses.
In particular, the rising part of the $dG_\text{I} / dV_l$ curve (grey in Fig.~\ref{fig:thermo}(b)) corresponds to the local minima of $G_\text{total}$ at different temperatures.
These are (meta-)stable equilibrium states, hence trace the equilibrium curve that describes the quasi-static process.
The maximum of $dG_\text{I} / dV_l$ corresponds to the upper temperature limit of the quasi-static melting, namely, the ``de facto'' melting temperature $T_\text{m}$ of the system.


\medskip

In contrast, a typical Gibbs-Thomson system does not have quasi-static melting.
Fig.~\ref{fig:thermo}(c) shows the total free energy as the function of $V_l$ for a small spherical solid particle (radius $r_0$) covered with liquid melt.
Again, three different temperatures are chosen: the bulk melting temperature $T_0$, the ``de facto'' melting temperature $T_m$, and a temperature lower than $T_m$.
It could be calculated that $dG_\text{I} / dV_l = -2\gamma_\text{sl} /(r_0^3 - 3V_l/(4\pi))^{-1/3}$, and has the maximum at $V_l=0$.
Hence there is no (meta-)stable equilibrium. The ``de facto'' melting temperature $T_\text{m}$ is characterized by an unstable equilibrium at $V_l=0$.


\subsection{Simulation details}
\begin{figure}
  \centering
  \includegraphics[width=\columnwidth]{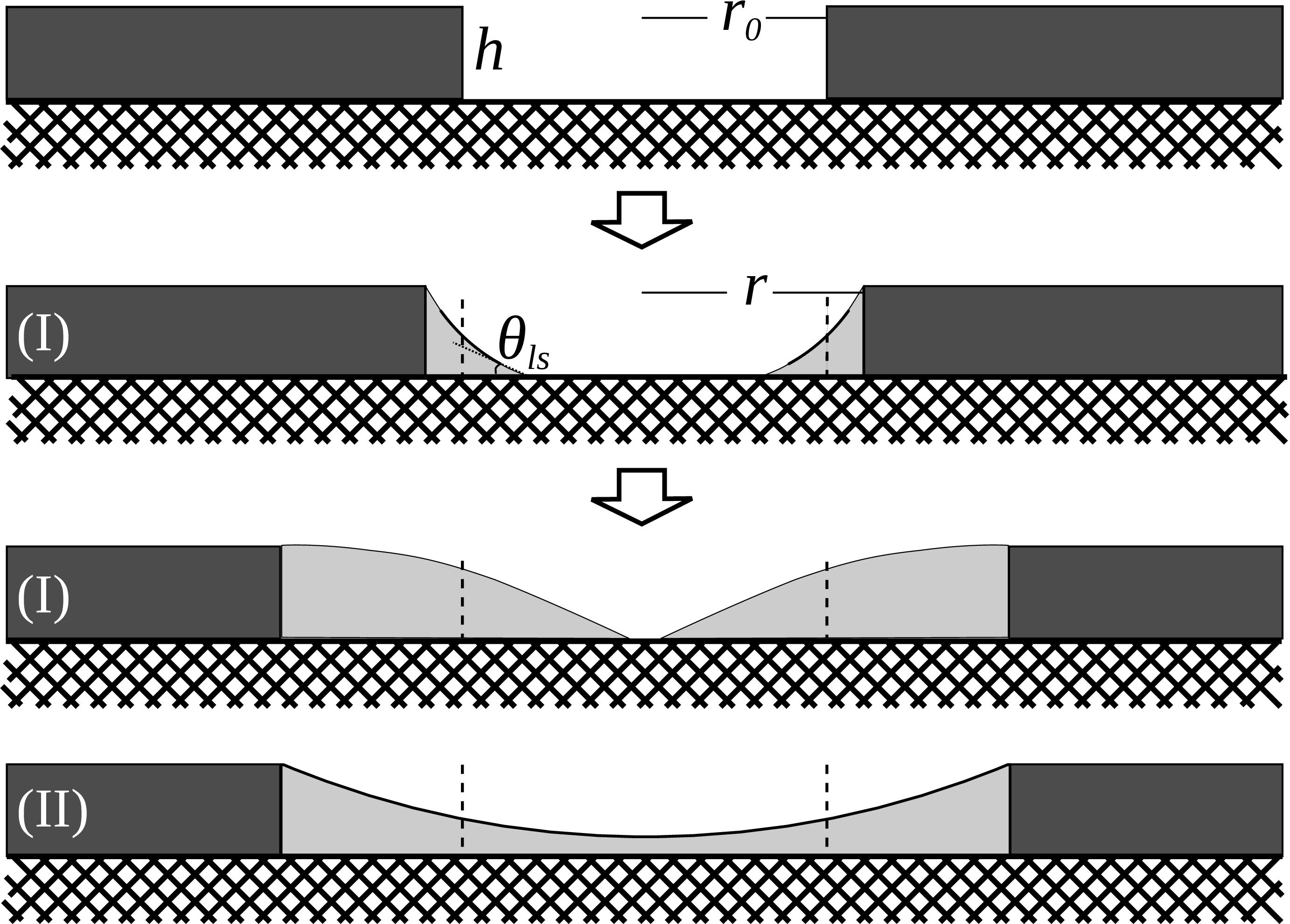}
  \caption{Melting process from a hole of finite size ($r_0$) in an infinite film of uniform thickness $h$. Drawings illustrate the cross-section. The liquid wets the lateral surface completely and the substrate partially with contact angle $\theta_{ls}$. Liquid melt first accumulate at the edge as ``open'' rim (I) , then the liquid front coalesce and form a ``closed'' concave spherical cap (II).}
 \label{fig:hole-scheme}
\end{figure}

We focus on the melting of a film between two different media: a substrate and air. Melting of a film between identical media or substrates is either straight-forward, or closely related. A cylindrical hole (radius $r_0$) in the film exposing the substrate would be the starting point of melting. We assume the followings:
\begin{itemize}
  \item The hole in the solid maintains a vertical cylinder geometry. The substrate is planar and the film is of uniform thickness, hence the cylindrical hole also maintains a constant height $h$ during melting;
  \item The total volume of solid and liquid together remains constant as the density difference between solid and liquid is neglected;
  \item The liquid wets completely the lateral surface of the hole, but only partially the film surface and the substrate, with a non-zero contact angle;
  \item The liquid is pinned at the solid/air/liquid contact line of the hole edge, in another word, there is no fixed contact angle at this line; 
  \item Young's equation holds for contact angles at the liquid/substrate/air contact lines. 
\end{itemize}

\begin{figure}
  \centering
  \includegraphics[width=\columnwidth]{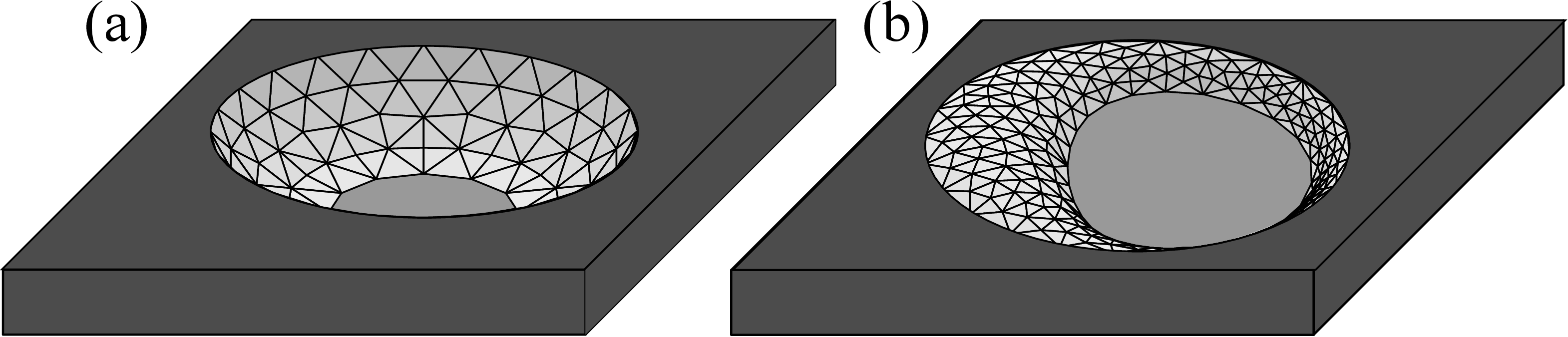}
  \caption{Two types of ``open''morphologies: (a) rouloid, or rotational-symmetric constant-mean-curvature rim and (b) bulged rim.}
  \label{fig:hole-scheme-bulge}
\end{figure}

The non-trivial shape of the capillary surface during the melting is solved by relaxing the surface in Surface Evolver \cite{brakke1992evolver} with a given $V_l$ and corresponding boundary conditions.
For the simulations we use real data of long-chain alkane systems \cite{Dirand2002enthalpy, yi2011MD, riegler2007pre}: liquid/air interfacial tension $\gamma_\text{lv}=25 \times 10^{-3}$N/m, $\gamma_\text{lw} \text{(hole side wall)} = 10 \times 10^{-3}$N/m, $\gamma_\text{ls}\text{(film surface and substrate)} = 4 \times 10^{-3}$N/m,  and $\Delta S_\text{fus} = 5 \times 10^5$ J/Km$^{-3}$.
The contact angle is $\theta_\text{ls}=15^\circ$ on both the film surface and the substrate.
The model can be easily tuned to simulate different systems.
For example, if $\theta_\text{ls}=90^\circ$, the model represents not only a film coated on a substrate, but also a self-supporting film in the air (after a reflection in the substrate plane).


Fig.~\ref{fig:hole-scheme} illustrates different liquid shapes that may emerge during melting.
The hole, from where the melting starts, has an initial radius $r_0$ and a depth $h$ (the film thickness), $r_0 >h$.
During the melting, when the radius of the hole becomes $r$, the liquid volume $V_l = \pi h (r^2-r_0^2)$.
Liquid melt will first appear as a symmetric rim at the edge of the hole.
More precisely, the capillary surface will be a constant-mean-curvature surface of revolution (Delaunay surface, or rouloid) \cite{delaunay1841surface, Eells1987surfaces}.
As the liquid volume increases, two different situations are possible: 
either there remains an opening exposing part of the substrate (``open'' shape, (I)), or the liquid covers the entire substrate (``closed'' shape (II)).

The ``closed'' shape has a geometric limit: the top surface of the liquid is a concave spherical cap whose boundary is pinned at the upper edge, hence the ``closed'' shape will only appear when a sufficient amount of the solid has melted. 
This limit can be calculated analytically. The radius of the hole has to be larger than a minimum radius $r_\text{min}$, which is: 
\begin{equation}
  \frac{r_\text{min}}{h} = \sqrt{2(\frac{r_0}{h})^2 -\frac{1}{3}}
  \label{eq:rmin}
\end{equation}

The ``open'' shape can be symmetric (rouloid) or non-symmetric (bulged); see Fig.~\ref{fig:hole-scheme-bulge}.
However, in this system, the bulged shape are only possible with sufficiently large $V_l$ that the system is already unstable, hence not the focus of this work.

In the following plots we use normalized, dimension-free variables: $\overline{G} = G/(\gamma_\text{lv} r_0 h)$, $\overline{V} = V/(\pi r_0^2 h)$, $\overline{r} = r/h$, $\overline{dG_\text{I} / dV_l} = dG_\text{I} / dV_l \cdot h/\gamma_\text{lv}$, and the in-plane curvature $\overline{\kappa} =\kappa \cdot h$.

\subsection{Experimental methods}
We use silicon wafer covered with a $300$ nm thick oxide layer as substrate. The oxide layer has a refractive index close to the long-chain alkane. Such a thickness is chosen because it provides interference enhancement of contrast from thin alkane layers on the substrate \cite{kohler2006optical}. 
We coat the piranha-cleaned wafer with long-chain alkane (triacontane \ce{C30H62} or hexatriacontane \ce{C36H74}) solution in toluene. After a heat-cool cycle, a ``surface frozen'' alkane monolayer \cite{Merkl1997SF} will be formed on the wafer. 
The sample is then annealed at a temperature slightly below $T_0$ for a few minutes. Films and islands of various size and height will appear. The suitable ones are chosen for the melting experiments.
The height of the alkane films are measured with atomic force microscope (AFM).

The substrate in such a system is actually the surface frozen alkane monolayer, covering the silica surface.
This layer, same with the top of alkane film, can only be partially wetted by liquid alkane melt with a contact angle around $15^o$. 
This contact angle is directly calculated from microscope observation or alkane droplets: The light (peak wavelength $550$ nm) reflected from liquid-air interface and the silica-silicon interface in the wafer interfere and form Newton's rings.
From this pattern we get the droplet surface profile and the contact angle.

\section{Discussions}
\subsection{Transitions between different morphologies during melting}

\begin{figure}
  \centering
  \includegraphics[width=\columnwidth]{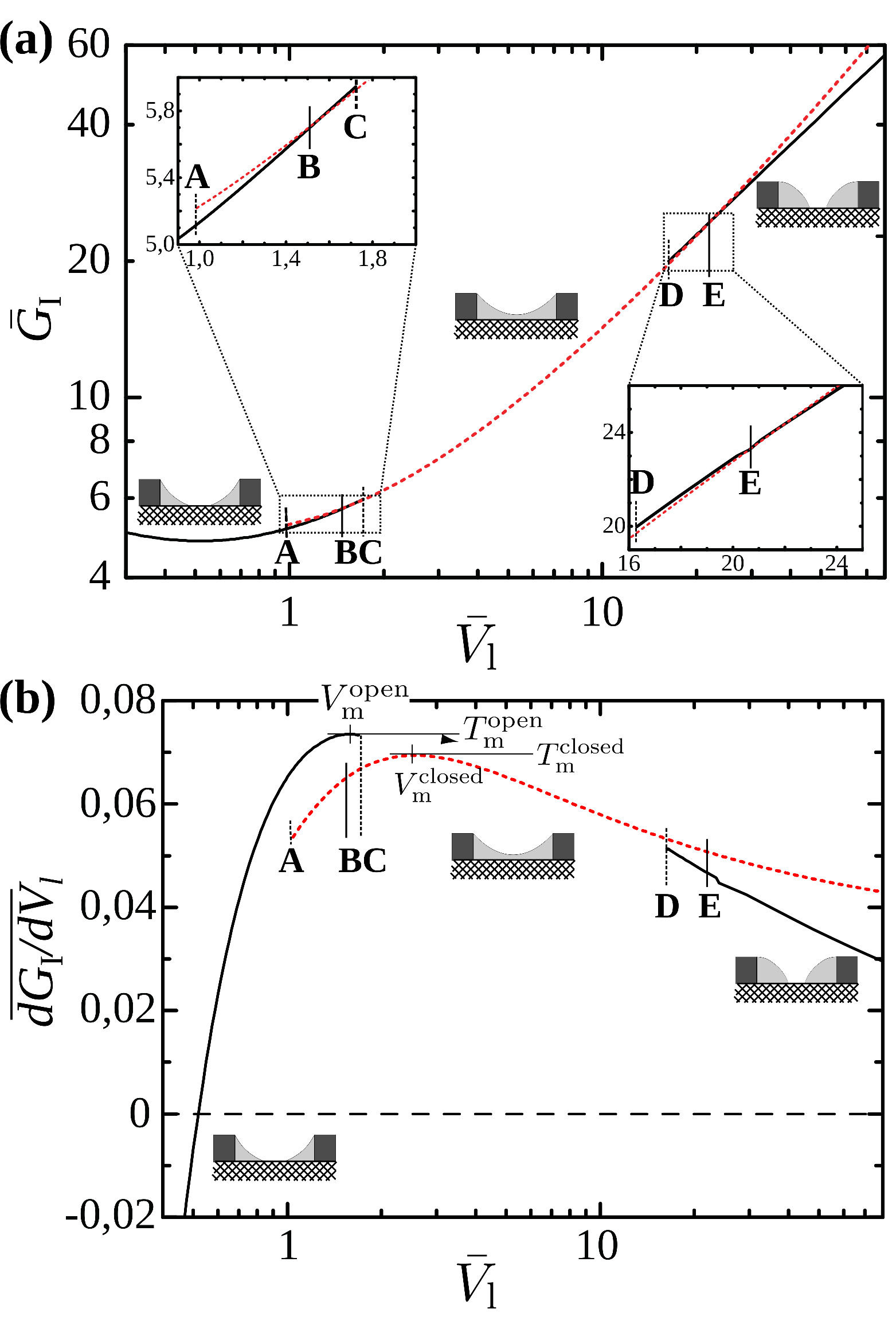}
  \caption{Melting process of a film from a hole of $\overline{r_0}=5$. (a) Normalized interfacial energy $\overline{G_\text{I}}$ as a function of normalized liquid volume $\overline{V_l}$. The solid black lines and the dashed red line indicate respectively the ``open'' and the ``closed'' shapes. For small volume (left of \textbf{A}) only ``open'' shape is possible. Between \textbf{C} and \textbf{D} only ``closed'' shape is possible. Between \textbf{A} and \textbf{C}, also above \textbf{D}, both shapes are possible but with a difference in energy, except when at \textbf{B} and \textbf{E}, the two iso-energy points. (b) The corresponding $\overline{dG_\text{I}/dV_l}$ as a function of $\overline{V_l}$.
}
  \label{fig:hole-energy}
\end{figure}

Fig.~\ref{fig:hole-energy} plots the normalized interfacial free energy $\overline{G_\text{I}}$ and $\overline{dG_\text{I}/dV_l}$ 
against the normalized liquid volume $\overline{V_l}$. The plot represents the melting process of a film from a hole of given size $\overline{r_0}=5$ on a substrate with $\theta_\text{ls} = 15^o$. 
This corresponds to the actual case of a hole about $500\,$nm in radius in an alkane film with $h=100\,$nm. 

Note that at the bulk melting temperature $T_0$, the $G_\text{I}$ plot also represents $G_\text{total}$. 
The eventual increase of $G_\text{total}$ indicates an incomplete melting. 
The minimum at a non-zero ${V_l}$ indicates a co-existence of liquid melt and the solid under equilibrium.
The same can be read from Fig.~\ref{fig:hole-energy}(b): As $\Delta \mu=0$ at $T_0$, the intersection of the rising part of the ${{dG_\text{I}}/{dV_l}}$ with the x-axis indicates a stable equilibrium.

By Eq.~\eqref{eq:rmin}, the minimal volume for the ``closed'' shape $\overline{V_\text{min}} \approx 1$, marked by \textbf{A}.
\textbf{B} and \textbf{E} mark the two points where the ``open'' and the ``closed'' shapes have the same energy.
Above \textbf{B}, the  ``closed'' shape becomes more favorable; above \textbf{E}, the ``open'' shape becomes more favorable.
\textbf{C} and \textbf{D} mark the two points between which the ``open'' shape is forbidden.

The melting scenario consists of three stages: (1) melting starts from a symmetric rim; (2) the liquid closes into a concave somewhere between \textbf{A} and \textbf{C}; (3) the ``closed'' liquid concave raptures into ``open'' rim somewhere after \textbf{D}. 
We will ignore the closed-to-open transition in step (3) as it only happens beyond the critical point of melting.
Because of an energy barrier between the two shapes, the open-to-closed transition can take place anywhere between \textbf{A} and \textbf{C}.

In Fig.~\ref{fig:hole-energy}(b), both the ``open'' and ``closed'' curves have maxima, corresponding to the upper limit of the quasi-static melting. The respective ``de facto'' melting points and the corresponding ``critical'' liquid volumes are denoted by ($T_\text{m}^\text{open}, \, V_\text{m}^\text{open}$ ) and ($T_\text{m}^\text{closed}, \, V_\text{m}^\text{closed}$ ).


\begin{figure}
  \centering
  \includegraphics[width=\columnwidth]{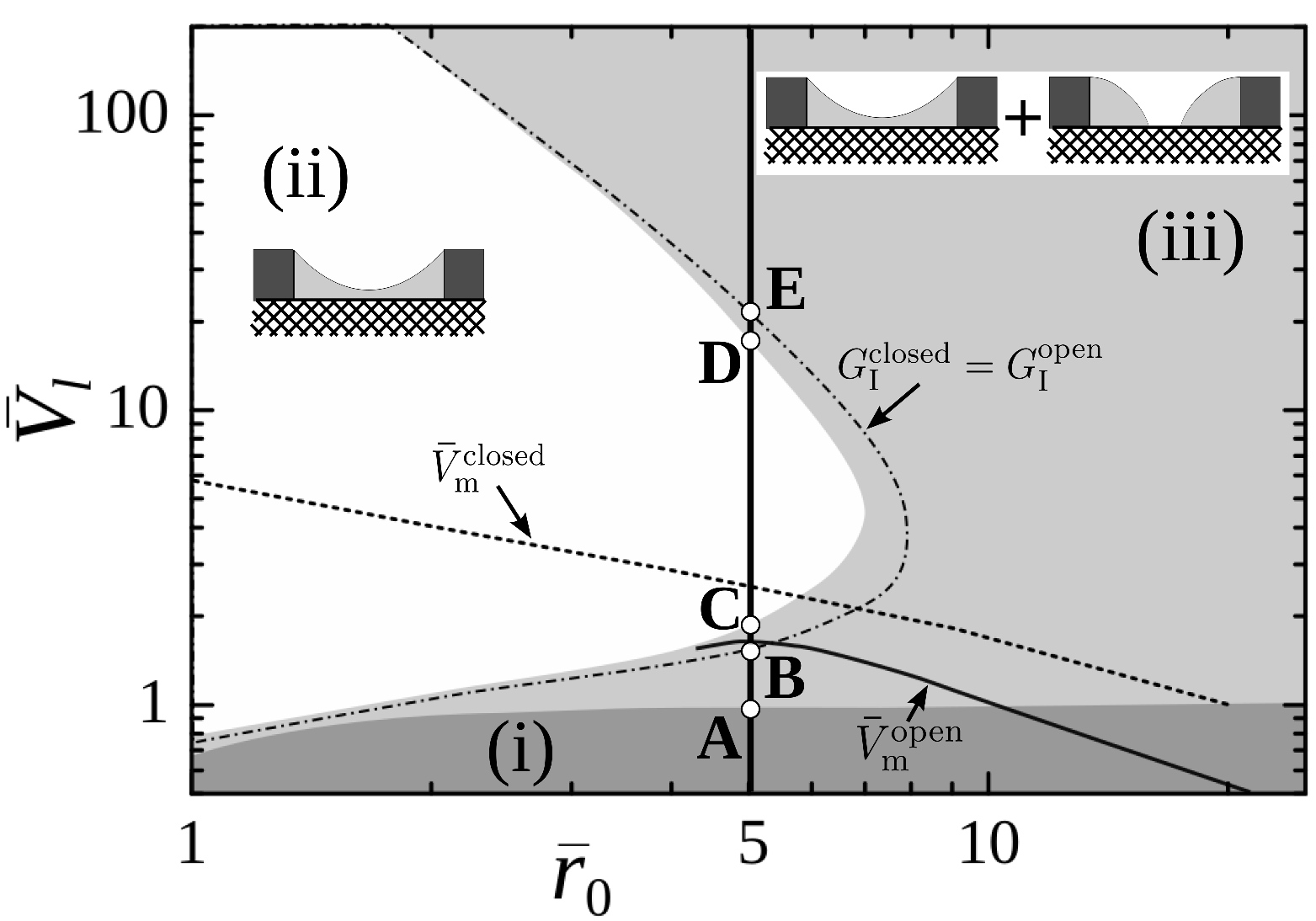}
  \caption{Phase diagram for the two different shapes (``open'' and ``closed'') parameterized by $\overline{V_l}$ and $\overline{r_0}$. The three regions colored differently show where (i) only ``open'' shape, (ii) only ``closed'' shape, and (iii) both shapes can exist due to the geometrical boundary conditions. Dash-dotted curve indicates where the ``open'' and ``closed'' shapes have the same interfacial energy. The solid and the dashed curves shows respectively the critical volumes of ``open'' and ``closed'' shape, i.e., the absolute volume limits of thermal stability. A volume-increasing path is indicated, with the points \textbf{A} to \textbf{E} same as in Fig.~\ref{fig:hole-energy}.
}
  \label{fig:hole-energy-phase}
\end{figure}

\medskip
The melting process of films from holes of different sizes is summarized in the phase diagram in Fig.~\ref{fig:hole-energy-phase},
parameterized by $\overline{V_l}$ and $\overline{r_0}$.
The phase diagram is divided into three regions where: (i) only ``open'' shape is possible (white), (ii) only ``closed'' shape is possible (dark grey) and (iii) both shapes are possible (light grey).
The boundary between the region (i) and region (iii) is derived analytically from Eq.~\eqref{eq:rmin}.
The dash-dotted curve is the iso-energy curve where ``open''  and ``closed'' shapes have the same interfacial energy ($G_\text{I}^\text{closed} = G_\text{I}^\text{open}$). 

The melting scenario presented in Fig.~\ref{fig:hole-energy}, with $\overline{r_0} =5$, corresponds to the vertical straight line, with the marks \textbf{A} to \textbf{E} bearing the same meanings.
We also plot the ``critical'' volumes $V_\text{m}^\text{open}$ (solid curve) and $V_\text{m}^\text{closed}$ (dotted curve), as indicated in Fig.~\ref{fig:hole-energy}b.


\subsection{Melting scenarios and blurred melting points}
\begin{figure}
  \centering
  \includegraphics[width=\columnwidth]{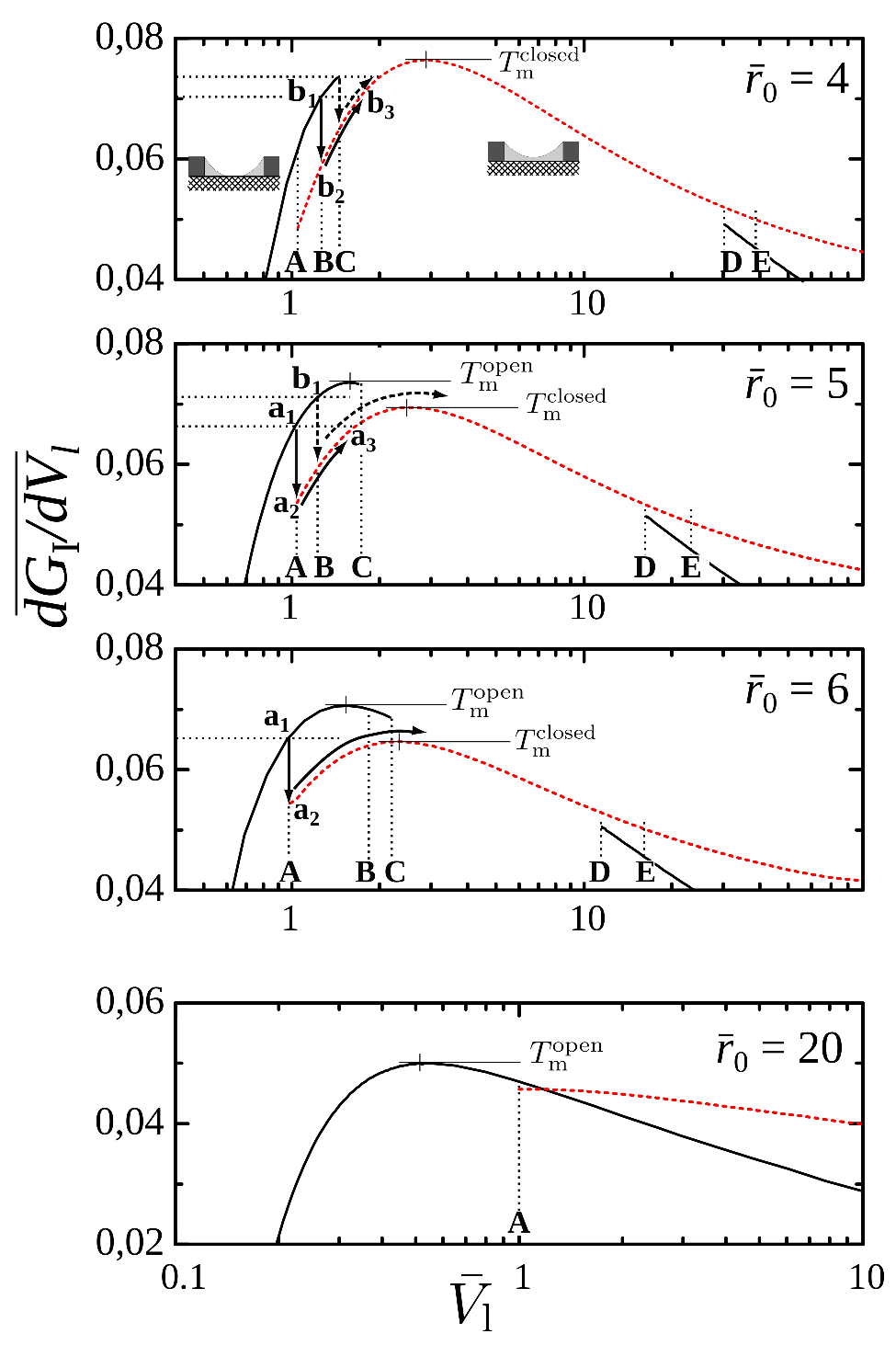}
  \caption{Melting scenarios of film from holes of different sizes. $\overline{dG_\text{I}/dV_l}$ is plotted as a function of $\overline{V_l}$. As discussed in the text, the rising parts are the equilibrium curves that describe the quasi-static melting processes.
  Same as in Fig.~\ref{fig:hole-energy}, the solid black lines and the dashed red lines indicate correspondingly the ``open'' rim and the ``closed'' concave. Point \textbf{A} to \textbf{E} have also the same meaning.
  $T_\text{m}^\text{open}$ and $T_\text{m}^\text{closed}$ indicate the ``de facto'' melting points of the corresponding morphology.}
    \label{fig:hole-dGi}
\end{figure}

In the following we look into the melting scenarios for different $\overline{r_0}$, i.e., the film melt from a hole of different given sizes. 
Fig.~\ref{fig:hole-dGi} shows the  $\overline{{dG_\text{I}}/{dV_l}}$ as a function of $\overline{V_l}$ with the same markings as in Fig.~\ref{fig:hole-energy}.
Note that for $\overline{r_0}=4, \, 5, \, 6$, both the ``open'' and the ``closed'' curves have rising parts.
Therefore, there are two equilibrium curves, each describes a quasi-static melting process.
At the overlapping part (\textbf{A}~--~\textbf{C}) the system can jump from one equilibrium curve to the other by overcoming an energy barrier.

For $\overline{r_0}=4$, melting start from liquid forming a symmetric rim at the hole edge.
Let us assume the morphology transition happens at \textbf{B}. From \textbf{b1} to \textbf{b2} it will first be an isochoric transition.
However, at this temperature, the corresponding liquid volume $V_l$ on the ``closed'' curve is much larger, hence $V_l$ will rapidly increase from \textbf{b2} to \textbf{b3}.
The system is still metastable because \textbf{b3} is not yet the maximum. 
Same analysis applies to any points between \textbf{A} and \textbf{C}; the system is always stable after transition. Hence the maxmium on the ``closed'' curve, $T_\text{m}^\text{closed}$, is the only ``de facto'' melting point in this system. 

In practice, this melting scenario allows us to seal small holes in the solid film: First rise the temperature slowly until there is enough liquid inside the hole to form a closed concave, while carefully keep the temperature below $T_\text{m}^\text{closed}$. Then quickly freeze the system and solidify the closed concave of liquid melt into a glass state. The hole is sealed.

For $\overline{r_0}=5$, the scenario becomes different around the open-to-close transition:
If the transition occurs at \textbf{A}, it goes from \textbf{a1} via \textbf{a2} to \textbf{a3}, after which the system is still metastable.
However, if the transition occurs at \textbf{B}, it will lead to a complete melting as the temperature at \textbf{b1} is already higher than $T_\text{m}^\text{closed}$.
It is also possible that the transition does not occur before the temperature reaches $T_\text{m}^\text{open}$, the maximum on the ``open'' curve.
In summary, the ``de facto'' melting point $T_\text{m}$ of this system is a temperature range rather than a single temperature.  
It can be any value between $T_\text{m}^\text{closed}$ and $T_\text{m}^\text{open}$, depending on when the morphology transition takes place.

For $\overline{r_0}=6$, even if the transition occurs at the minimum volume (\textbf{A}), the temperature is already higher than $T_\text{m}^\text{closed}$, and the system melts immediately.
The ``de facto'' melting point $T_\text{m}$ of this system is also a temperature range, but between $T_\textbf{a1}$ and $T_\text{m}^\text{open}$

In the case with $\overline{r_0}=20$, the liquid in the system remains a symmetric rim up to $T_\text{m}^\text{open}$.
``Closed'' shape is only possible after that, hence absolutely unstable in the sense of melting.

\subsection{Experimental results of different hole melting scenarios }

 \begin{figure}
   \centering
   \includegraphics[width=\columnwidth]{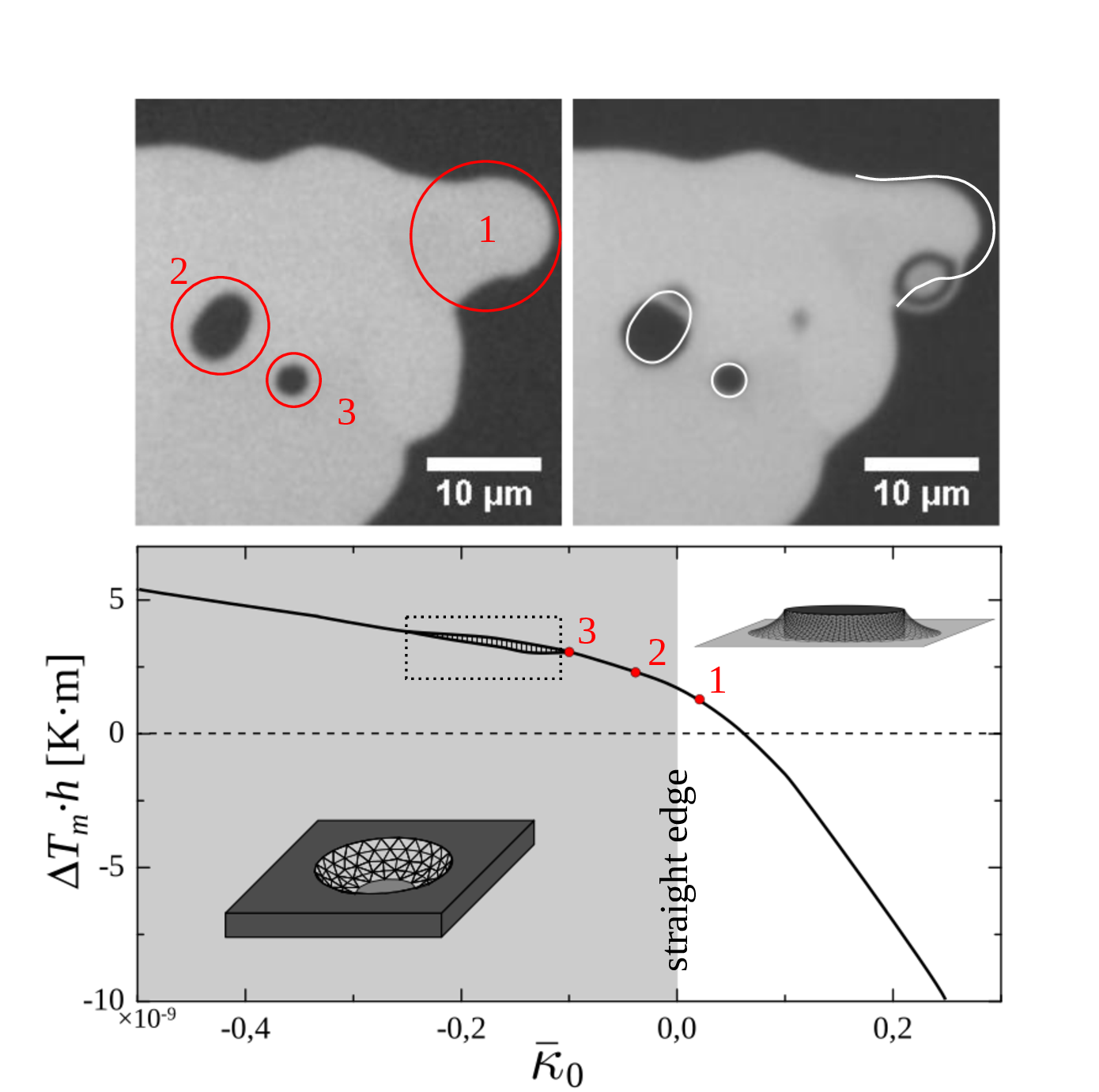}
   \caption{Top: Melting behavior of solid alkane film ($h\approx 100\,$nm) with three different local structures: (1) a protruding ``peninsula'' ($\overline{r_0} \approx 40, \, \overline{\kappa_0} \approx 0.025$); (2) a hole with $\overline{r_0}\approx 30, \, \overline{\kappa_0} \approx -0.033$; (3) a hole with $\overline{r_0} \approx 10, \, \overline{\kappa_0} \approx -0.1$. In the same experiment they melt sequentially. 
   Bottom: Simulated melting point shift (scaled to the height) of different geometries as a function of initial in-plane curvature (normalized). The model is based on the experimental system of alkane film / island melting. The grey region shows film melting from hole, and the white region shows melting of island. The boundary between the two region indicates melting of a straight terrace edge. Melting point shifts for the three local structures shown in top frame are marked respectively. Dotted rectangle highlights the geometries with ``de facto'' melting points being a range. Bulging, which happens during melting a less-curved structure and lowers $T_\text{m}$, is not shown on this plot. } 
 \label{fig:hole-exp}
 \end{figure}

Combined with our previous calculation on melting of cylindrical island \cite{jin2016island} and straight-edge terrace \cite{Halim2012bulge}, we have now the full picture of melting process of different geometries. In Fig.~\ref{fig:hole-exp} (bottom) we plot the scaled melting point shifts $\Delta T \cdot h$ in our model system as the function of the normalized in-plane curvature of the initial solid edge. 
We assign a positive in-plane curvature to the island, $\overline{\kappa_0} = h/ r_0$, and negative to the holes, $\overline{\kappa_0} = -h/ r_0$.
One sees the $\Delta T_\text{m}$ decreases monotonically with the curvature. As a consequence, if a film with irregular edge and holes melts, the melting will start first from the island (peninsula), then the straight edge, and lastly the tiny defects (holes) inside. 
This counter-intuitive sequence of melting is verified by the following experiment.


Fig.~\ref{fig:hole-exp} (top) shows the melting process of an alkane film (\ce{C36H74}) of ca.\ $100\,$nm thick.
The curved edges behaves locally like islands, and inside the film there are two holes of different sizes.
Top right panel shows how much solid is melt at different positions under the same temperature. At the ``peninsula'' (1), comparing to the solid boundary before melting (white line), large amount of solid disappears. The liquid accumulates as a big bulge. Around the big hole (2) some solid melt as well, while around the small hole (3) barely any change can be observed.
Such a melting sequence is consistent with the estimated ``de facto'' melting points at these three geometries: the ``peninsula'' (1) with the lowest ``de facto'' melting point melts the first, the small hole (3) with the highest ``de facto'' melting point melts the last.

Note that this sequence of melting makes it very hard to experimentally measure the melting point shift at a hole. 
Before the melting starts around the hole, it would have already started from the outer boundary.
Then, according to our previous study, the liquid melt would appear as drops that run and ``eat'' into the film at high speed \cite{Lazar2005movingdrop, Halim2012bulge}.
Hence the geometry would have been destroyed before any meaningful measurement.

\section{Conclusions}
When the contribution of the capillary surface (liquid/air interface) is taken into consideration, the melting process becomes non-trivial. 
We present a general approach for analysing these system.
More specifically, we first compute the energy of all the possible configurations with relaxed capillary surfaces, then deduce quasi-static process from the plot of energy derivative.

We apply this approach to thin films with holes of different sizes. We come to interesting conclusions, including an elevated ``de facto'' melting point that in some cases becomes a range, and a counter-intuitive sequence of melting that starts from outer boundary instead of defects inside. These are verified by experiments on thin films of long-chain alkanes between silica and air.

\section{Acknowledgement}
Discussion with H. M\"ohwald, H. Kusumaatmaja and H. Chen are gratefully acknowledged. C.Jin was support from IMPRS on Biomimetic Systems.

\bibliography{ref-hole.bib}

\end{document}